\documentclass[twocolumn,showpacs,preprintnumbers,superscriptaddress,amsmath,amssymb]{revtex4}
\usepackage{graphicx}
\usepackage{dcolumn}
\usepackage{bm}
\usepackage[utf8]{inputenc}
\usepackage{amsmath}
\usepackage{textcomp}
\usepackage{color}
\usepackage{amsfonts}%

\begin{document}

\title{Anomalous velocity distributions in active Brownian suspensions}

\author{Andrea Fiege}

\affiliation{Georg-August-Universit\"at G\"ottingen,  Institut f\"ur 
  Theoretische Physik, Friedrich-Hund-Platz 1, 37077 G\"ottingen,
  Germany}
\author{Benjamin Vollmayr-Lee}

\affiliation{Department of Physics and Astronomy, Bucknell University,
      Lewisburg, Pennsylvania 17837, USA}
\author{Annette Zippelius}
\affiliation{Georg-August-Universit\"at G\"ottingen,  Institut f\"ur 
  Theoretische Physik, Friedrich-Hund-Platz 1, 37077 G\"ottingen,
  Germany}
\affiliation{Max-Planck-Institut f\"ur Dynamik und Selbstorganisation,
  Am Fa\ss berg 17, 37077 G\"ottingen, Germany}

\date{\today}

\begin{abstract}
  Large scale simulations and analytical theory have been combined to
  obtain the non-equilibrium velocity distribution, $f(v)$, of
  randomly accelerated particles in suspension. The simulations are
  based on an event-driven algorithm, generalised to include
  friction. They reveal strongly anomalous but largely universal
  distributions which are independent of volume fraction and collision
  processes, which suggests a one-particle model should capture all
  the essential features. We have formulated this one-particle model
  and solved it analytically in the limit of strong damping, where we
  find that $f(v)$ decays as $1/v$ for multiple decades, eventually
  crossing over to a Gaussian decay for the largest velocities. Many
  particle simulations and numerical solution of the one-particle
  model agree for all values of the damping. 
\end{abstract}



\maketitle

\section{Introduction}

In recent years there has been growing interest
in so called active matter, referring to the ability of the
constituents to move actively by either extracting energy from the
environment or depleting an internal energy depot. Examples are motor
proteins, bacterial swimmers or motile cells~\cite{Lauga2009}. Whereas
the mechanism that drives the individual active particle has been
studied for many years~\cite{Lighthill52,Brennen1977,Golestanian2008},
the collective behavior of a large number of individuals has been
addressed only recently. Very rich behavior has been observed,
ranging from pattern formation and nonequilibrium phase transitions to
turbulence~\cite{Baskaran2009,Loewen2012}. Active particles on
mesoscopic to macroscopic scales have also been realized in the
form of self-propelled colloids (Janus particles)~\cite{Suzuki2011}
and vibrated polar granular rods~\cite{Kudrolli2008}. More generally,
granular particles that are driven by random kicks may be
considered active matter with, however, the direction of motion being
random.

Our focus here are the velocity distributions of active
particles in suspension. Whereas in equilibrium, the velocities
universally follow the Maxwell-Boltzmann distribution, this does not
hold for nonequilibrium stationary states, where in general deviations
from the Maxwell-Boltzmann distribution are observed. Few studies have
focused on the velocity distribution in the context of active cell and
bacteria suspensions \cite{czirok1998exponential,sokolov2010swimming}. 
In \cite{czirok1998exponential} extensive experimental data were taken
for several cell types, allowing for a statistical analysis of the
cell's velocities. The authors concluded that exponential
distributions are a general characteristic feature of cell
motility. Such exponential distributions have indeed been found in
models of active Brownian particles~\cite{Romanczuk2012}; however
other distributions have been seen as well, depending on the mechanism
of self-propulsion~\cite{Romanczuk2012,llopis2007dynamic}.

For driven granular media on the other side, numerous studies have
been performed to analyze velocity distributions. In experiments,
various driving mechanisms were shown to produce non-Gaussian velocity
distributions~\cite{abbas2006dynamics,Urbach2005,PhysRevE.62.R1489,PhysRevE.70.040301,Reis2007,Menon2009}. If
the particle's motion is strongly damped either due to the surrounding
fluid or due to collisions with the wall, the velocity distributions
are exponential. In~\cite{PhysRevE.70.040301} the authors use a
single-particle simulation of a frictional particle to explain the
observed velocity distribution. Their argument was turned into a
Fokker-Planck equation~\cite{BenNaim2005,harting2008anomalous}, whose
stationary solution is in good agreement with
experiment~\cite{BenNaim2005}. 

In the present work we study a simple model of active particles in a
suspension, described below, using event-driven simulations. We
obtain nearly universal velocity
distributions which depend primarily on a single parameter, and
which exhibit significant deviations from Gaussian behavior, but also
non-exponential tails (see Fig.~\ref{fig:velocity_dist_0.35}). Further, we
develop a single-particle theory that shows good agreement with the
simulation data.

\section{Model}

Here we discuss a simple model for active
particles: hard spheres placed in a fluid with a viscous drag $\gamma$,
that are accelerated at discrete times and undergo elastic collisions.

The equation of motion for particle $i$ reads
\begin{equation}
  \partial_t\mathbf  v_i = -\gamma\mathbf  v_i + \left. \frac{\Delta\mathbf  v_i}{\Delta t} 
  \right\vert_{coll} + \left. \frac{\Delta \mathbf v_i}{\Delta t} 
  \right\vert_{Dr}. \label{eq:langevin}
\end{equation}
The driving force is modeled as discrete kicks with amplitude $\Delta
\mathbf p = m \Delta \mathbf v$ and frequency $f_{Dr}$. The
components of the kick velocity, e.g.
$\Delta v_x$, are drawn from a Gaussian distribution with mean $0$ and
variance $\sigma^2$:
\begin{equation}
P(\Delta v_x)=\frac{1}{\sqrt{2\pi}\sigma}\exp\left(-\frac{(\Delta v_x)^2}{2\sigma^2}\right)
\label{eq:kick_distribution}
\end{equation}
and for the other components accordingly. We ignore hydrodynamic
interactions. 

The above dynamics is a very crude approximation to the run-and-tumble
behavior of bacteria, such as {\it E.Coli} and
others~\cite{Berg2004,Cates2008,Golestanian2012}. In a time interval
$\Delta t$, a particle is accelerated once and subsequently performs a
random motion determined by the surrounding fluid and interactions
with the other particles. If the bacteria acceleration events
(strokes) are sufficiently rare, subsequent kicks may be regarded as
uncorrelated in direction, so that our model should apply. 

We are interested in a steady state, where the energy due to
dissipation is balanced by the energy input due to random kicks:
\begin{equation}
2m\gamma \langle v^2\rangle = d \, f_{Dr} m \sigma^2
\end{equation}
where $d$ is the dimensionality of the system. In the following we
will choose units such that lengths are measured in units of particle
radius and mass in units of particle mass. We choose the time scale
so that the average steady state kinetic energy is $d/2$, which
corresponds to $k_BT=1$ for a thermal system. In these units the
driving amplitude becomes $\sigma^2=2\gamma/f_{Dr}$, leaving three
independent parameters: $\gamma$, $f_{Dr}$, and the volume fraction
$\eta$. We will consider moderately dilute systems for which the
particle collision frequency is well-described by the Enskog result
$\omega_{coll}(\eta) = 12\chi\eta/\sqrt{\pi}$ with the Carnahan-Starling
expression for the pair correlation at contact $\chi =
(1-\eta/2)/(1-\eta)^3$. Thus our three parameters provide three
independent time scales: $\gamma$, $f_{Dr}$, and $\omega_{coll}$
(in place of $\eta$).

\section{Simulations}

We performed event driven simulations of hard
spheres. The original algorithm \cite{alder1959,Lubachevsky1991} was
changed in order to implement friction as in \cite{fiege2012dynamics}.
The main effort of an event-driven simulation of ballistically moving
particles goes into the calculation whether two particles will collide
or not. If a collision between particle $i$ and $j$ will occur, the
difference of their trajectories,
\begin{equation}
 \mathbf r_i(t) - \mathbf r_j(t)\equiv
  \mathbf r_{i,j}(t)  =   \mathbf r_{i,j} (t_{0}) +  
\mathbf v_{i,j}(t_{0})(t-t_0)
\end{equation}
must be equal to the sum of their radii at time $t_{coll}$, i.e.,
\begin{equation}
 R_i + R_j = \left\vert  \mathbf r_{i,j}(t_{coll})  \right\vert
\end{equation}
yielding a quadratic equation in $t_{coll}-t_0$. For the damped
motion, $\gamma\neq 0 $, one can still integrate the equations of
motion in between collisions analytically:
\begin{equation}
  \mathbf r_{i,j}(t)  =   \mathbf r_{i,j} (t_{0}) +  
\mathbf v_{i,j}(t_{0})\frac{1 - e^{- \gamma(t-t_{0})}}{\gamma} 
\label{equ:damped_motion}
\end{equation}
Compared to ballistic motion, the linear time interval between two
collisions $(t_{coll}-t_0)$ is simply replaced by $(1 - e^{-
  \gamma(t_{coll}-t_{0})})/\gamma$. Since we know the collision time
from the ballistic simulation, we can just use the above relation to
determine the collision times for the damped system. The remaining
events in the simulation---driving events, wall collisions, sub-box
wall collisions---are handled accordingly. The only remaining
difference in the damped system is that the place of a collision with
another particle or a (sub-box) wall might not be within range of the
damped motion. If this is the case, the collision will not occur,
instead the particle will slow down until a driving event takes place.

We have simulated a 3-dimensional system of 2122416 monodisperse
spheres with volume fractions $\eta = 0.05$ and $0.35$, corresponding to $\omega_{coll}=0.385$ and $\omega_{coll}=7.11$. The system is
equilibrated with $\gamma=0$ and no forcing. Subsequently, damping and
the acceleration force are switched on. Then, after another $100$
collisions per particle to ensure relaxation to a stationary state,
the velocity distribution is measured. These simulations were
conducted for various values of the parameters $\gamma$, $f_{Dr}$, and
$\eta$.

\begin{figure}
\includegraphics[width=.49\textwidth]{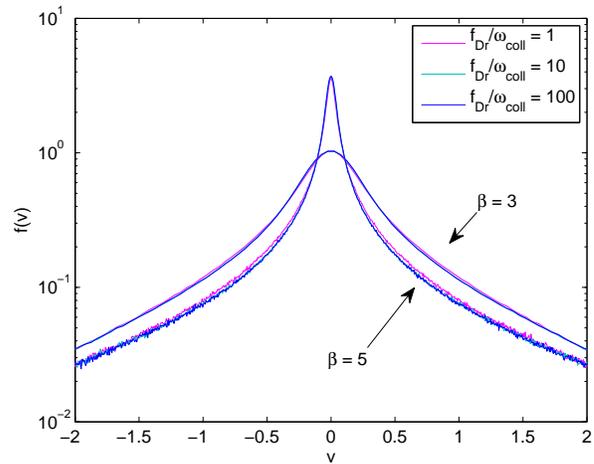}
\caption{(colour on-line) Testing the dependence of the velocity
distribution on the ratio  $f_{Dr}/\omega_{coll}$ for two
values of $\beta$. The volume fraction is taken to be 
$\eta=0.35$ corresponding to $\omega_{coll}=7.11$.}
\label{fig:velocity_dist_coll}
\end{figure}

For most simulations we set the driving frequency equal to the Enskog
collision frequency, $f_{Dr}=\omega_{coll}$, except for
Fig.~\ref{fig:velocity_dist_coll} where we explicitly study the
effects of changing the driving frequency. 
The observed collision frequencies match the Enskog expression for
small damping and decrease for larger damping by at most $40\%$ for
the largest damping considered here. Hence our choice,
$f_{Dr}=\omega_{coll}$, implies that typically
a particle collides once before it is kicked again.

\begin{figure}
\includegraphics[width=.49\textwidth]{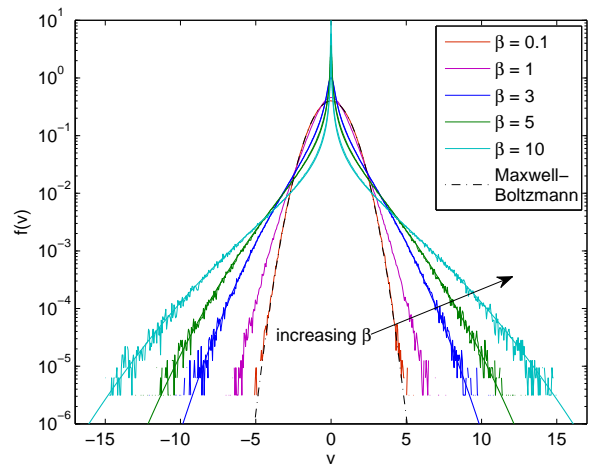}
\caption{(colour on-line) Velocity distributions for volume fraction
  $\eta=0.35$, $f_{Dr}=\omega_{coll}=7.11$, and several values of
  $\beta=\gamma/f_{Dr} = 0.1,1,3,5,10$. The dashed-dotted line shows the
  Maxwell-Boltzmann distribution. The coloured solid lines show the
  first iterative solutions of the one-particle model (see text below)
  for $\beta = 3,5,10$.}
\label{fig:velocity_dist_0.35}
\end{figure}

In Fig.~\ref{fig:velocity_dist_0.35} we show the velocity distribution
for $\eta=0.35$ and $f_{Dr}=\omega_{coll}$
 with various values of the damping constant
$\gamma$. The curves are labeled by the ratio
$\beta:=\gamma/f_{Dr}$. Whereas for very small $\beta$ the
distribution is approximately Gaussian, we observe increasingly strong
deviations for larger $\beta$. Small velocities are highly
overpopulated with an indication of a singularity in the limit of
large $\beta$. High velocities are overpopulated as well as compared
to the equilibrium Maxwell-Boltzmann distribution. These deviations
can be understood intuitively as follows: particles which have not
been recently kicked are damped to nearly zero velocity, whereas the
recently kicked particles populate the tail.

\begin{figure}
\includegraphics[width=.49\textwidth]{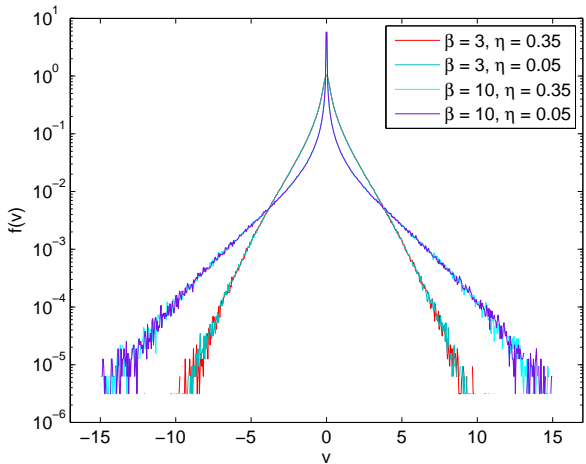}
\caption{(colour on-line) Testing the dependence of the velocity distribution
  on the volume fraction $\eta$. Data for $\eta=0.05$ and $\eta=0.35$
are shown for $\beta=3$ and for $\beta=10$. 
In both cases we find little to no dependence on the volume fraction.
The driving frequency is taken to be $f_{Dr}=\omega_{coll}$.}
\label{fig:velocity_dist_0.05}
\end{figure}

Next, we demonstrate the universality of these distributions. The
three-dimensional parameter space can be spanned by the parameters
$\eta$, $\beta$, and $f_{Dr}$. In
Fig.~\ref{fig:velocity_dist_0.05} we test the volume fraction
dependence of the velocity distribution. We set
$f_{Dr}/\omega_{coll}=1$ and compare for a given value of $\beta$,
e.g., $\beta=3$, the velocity distributions for two volume fractions,
$\eta=0.05$ and $\eta=0.35$, and find no discernable difference
between the distributions. This holds for all investigated values of
$\beta$, and is shown in Fig.~\ref{fig:velocity_dist_0.05} for
$\beta=3$ and $10$.


Then, in Fig.~\ref{fig:velocity_dist_coll} we test the dependence of
the velocity distribution on the ratio $f_{Dr}/\omega_{coll}$. Data
for $f_{Dr}/\omega_{coll}=1$, $10$, and $100$ are shown for $\beta=3$
and for $\beta=5$. For a specific value of $\beta$, the curves for
different $f_{Dr}/\omega_{coll}$ lie essentially on top of each other.

We summarize the main results of our simulations:
\begin{itemize}
\item{The distribution is independent of volume
  fraction for the investigated
  range of $\eta$.}
\item{It is also independent of the ratio $f_{Dr}/\omega_{coll}$; we
    obtain the same distribution, no matter whether a particle is
    kicked once or a 100 times in between collisions.}
\item{Consequently the distribution is almost exclusively determined
    by the ratio $\beta=\gamma/f_{Dr}$, even though the model contains
    three independent time scales, $\gamma$, $f_{Dr}$ and
    $\omega_{coll}$.}
\item{The one particle velocity distribution is Gaussian only in the
  limit $\beta\to 0$}. The distribution shows increasingly stronger
  deviations at {\bf {small}} and {\bf {large}} velocities for
  increasing $\beta$.
\item{The distribution seems to develop a singularity at small
  argument as $\beta\to\infty$. }
\end{itemize}
These observations, in particular the insensitivity to collision rate,
have led us to derive an approximate analytical theory for the
velocity distribution based on a single-particle model that neglects
collisions.

\section{Single-Particle-Model}

For simplicity, we consider one spatial
dimension only, assuming that the cartesian components of the velocity
are independent. With $\Delta t=1/f_{Dr}$, we consider the time
interval $[0,\Delta t)$, within which each particle gets one velocity
  kick at some random time. The idea of the calculation is the
  following: we use the one particle distribution at the beginning of
  the interval as input and compute the resulting one particle
  distribution at end of the interval, and then 
  require the two distributions to be the same in the stationary
  state. The speed of a single particle decreases in $\Delta t$ due to
  damping and generally increases due to a velocity kick, denoted by
  $\Delta v$. The kick occurs at time $\tau $ with probability
  $w(\tau)=\frac{1}{\Delta t}$ provided $ 0\leq \tau \leq \Delta
  t$. We are interested in the velocity distribution at the end of the
  time interval, when the kick velocity has decayed to $\Delta
  v_f=\Delta v\exp(-\gamma (\Delta t-\tau))$. For a given (fixed) kick
  size $\Delta v$, this quantity is a random variable due to the
  stochastic occurrence of the kick in the given time interval. The
  conditional probability to find a velocity $\Delta v_f$ for a given
  kick size $\Delta v$ is easily computed from the distribution
  $w(\tau)$:
\begin{equation}
p_k(\Delta v_f\vert \Delta v) = 
\begin{cases}
\displaystyle\frac{1}{\beta }\frac{1}{|\Delta v_f|}, & e^{-\beta} \leq 
  \Delta v_f/\Delta v\leq 1 \\[2ex] 
0 & \text{else}%
\end{cases}%
\label{eq:cond_prob}
\end{equation}
To obtain the non-conditional probability, $p_k(\Delta v_f)$ we write
\begin{align}
p_k(\Delta v_f) &= \int_ {-\infty}^\infty d\Delta v\; p_k(\Delta
v_f\vert 
\Delta v) \; P(\Delta v) \nonumber\\
&=\frac{1}{\beta }\frac{1}{\Delta v_f}    
\int_{\Delta v_f}^{\Delta v_f e^\beta} d\Delta v \cdot P(\Delta v) 
\label{eq:pk}
\end{align}
where $P(\Delta v)$ is the probability distribution for the kick
velocity, given by Eq.~(\ref{eq:kick_distribution}) with standard
deviation $\sigma=\sqrt{2\beta}$.

The total velocity at the end of the time interval, $v_f=\Delta
v_f+\tilde{v}$ is the sum of two terms: the kick velocity and the
velocity from the start of the interval, $v_i$, propagated in time to
the end of the interval, $\tilde{v}=v_i e^{-\beta}$. Given the
distribution of the initial velocities $f_i(v_i)$, the distribution of
final velocities (without kick) is given by
$\tilde{f}(\tilde{v})=f_i(\tilde{v}e^{\beta})e^{\beta}$. Since the two
velocity contributions $\Delta v_f$ and $\tilde{v}$ are statistically
independent, the probability distribution of the sum is given by the
convolution: $f(v_f)=(p_k*\tilde{f})(v_f)$. In the stationary state,
we require that the initial velocity distribution is equal to the
final velocity distribution, 
\begin{equation}
\label{eq:defining_eq}
f(v)=\int_{-\infty}^{\infty} du \; p_k(v-u) \; f(u e^{\beta}) \; e^{\beta}.
\end{equation}
The probability distribution within this single-particle model is a
function of $\beta=\gamma/f_{Dr}$ only, which matches the behavior of
the many-particle simulation data. With the Fourier transform $\hat
f(k)\equiv \int dv\, e^{ikv} f(v)$, the above equation simplifies,
\begin{equation}
\hat f(k) = \hat p_k(k) \hat f(ke^{-\beta}),
\end{equation}
and is solved by
\begin{equation}
\label{eq:fhat_soln}
\hat f(k) = \prod_{j=0}^\infty \hat p_k(k e^{-j\beta})
\end{equation}
with
\begin{equation}
 \hat p_k(k) = \int_0^1 dw \; \exp\left(-\frac{1}{2} k^2 \sigma^2
 e^{-2\beta w}\right).
\end{equation}
For a given $\beta$, the infinite product can be truncated for
some value of $j\gg 1/\beta$.

We now analyze the behavior of this formal solution,
Eq.~(\ref{eq:fhat_soln}), in the limits of large and small $\beta$,
where we can obtain simple analytic expressions for $f(v)$, and for
intermediate values of $\beta$, where we obtain the distribution
through an iterative numerical method.

First, in the $\beta\to 0$ limit, the $\Delta v_f$ distribution
$p_k(\Delta v_f)$ goes to $P(\Delta v_f)$, which is a Gaussian. Thus,
according to Eq.~(\ref{eq:defining_eq}), the velocity distribution
$f(v)$ must map to itself under a convolution with a Gaussian, which
requires that $f(v)$ must itself be a Gaussian. This stationary limit
corresponds to a continuous Ornstein-Uhlenbeck process
\cite{van1992stochastic}. The cumulant relation \cite{endnote}
implied by Eq.~(\ref{eq:fhat_soln}) specifies that the variance of
$f(v)$ goes to unity as $\beta\to 0$. This is confirmed by the
simulations with $\beta = 0.1$, shown in
Fig.~\ref{fig:velocity_dist_0.35}.

\begin{figure}
\includegraphics[width=.49\textwidth]{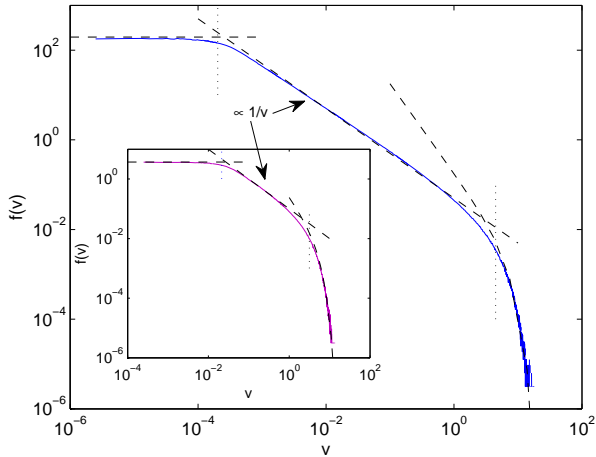}
\caption{(colour on-line) The three asymptotic solutions from Eq.~(\ref{eq:large_beta}) (dashed lines) and simulation data for $\beta =10$, and $\beta =5$ in the inset (both simulation data for $\eta=0.35$). The dotted lines depict the range limits from Eq.~(\ref{eq:large_beta}). The dashed lines are the analytic results from Eq.~(\ref{eq:large_beta}), without any fitting.}
  \label{fig:beta5_asympt}
\end{figure}

Second, in the large $\beta$ limit, only the $j=0$ term in
Eq.~(\ref{eq:fhat_soln}) contributes to the product, and so $f(v) =
p_k(v)$. As such, from
Eq.~(\ref{eq:pk}), we can identify three regions:
\begin{equation}
f(v) \approx
\begin{cases}
\displaystyle \frac{e^\beta - 1}{2\sqrt{\pi\beta^3}} & |v| \ll \sigma e^{-\beta} \\[2ex] 
\displaystyle \frac{1}{2\beta|v|} & \sigma e^{-\beta} \ll |v| \ll \sigma  \\[2ex]
\displaystyle \frac{1}{\sqrt{\pi\beta}}\frac{1}{v^2}e^{-v^2/4\beta} & |v| \gg \sigma 
\end{cases}%
\label{eq:large_beta}
\end{equation}
The middle case corresponds to taking the integration range in
Eq.~(\ref{eq:pk}) to be zero to infinity (for positive $\Delta v_f$).
The top case corresponds to a smooth cutoff to the $1/|\Delta v_f|$
behavior as $\Delta v_f\to 0$. The large $\Delta v_f$ limit is
obtained by setting the upper integration limit to infinity, giving
the complementary error function. In Fig. \ref{fig:beta5_asympt} we
plot the velocity distribution data for $\beta=10$ and $\beta=5$ and
compare to the analytic expressions (dashed lines) from
Eq.~(\ref{eq:large_beta}) and their ranges (dotted lines). The three
regions are clearly distinguishable and match the simulation data
well. Note that the $1/|v|$ region shrinks as $\beta$ decreases.

\begin{figure}
\includegraphics[width=.49\textwidth]{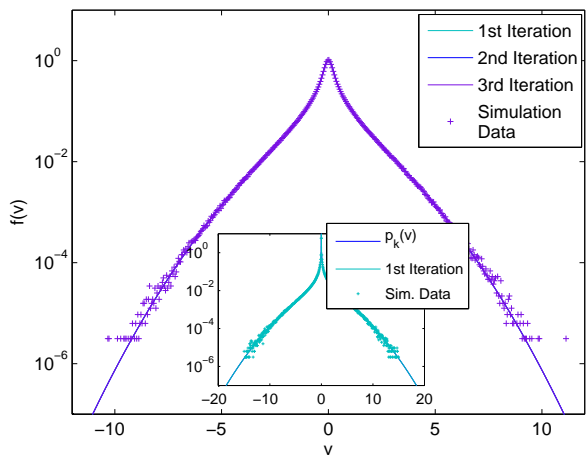}
\caption{(colour on-line) Main part: The first three iterations for
  $\beta = 3$ are almost indistinguishable and agree with the
  simulation data. Inset: First iteration and simulation data for
  $\beta = 10$, and $p_k(v)$ which is indistinguishable from the first
  iteration.}
  \label{fig:beta5}
\end{figure}

Third, for intermediate values of $\beta$ we solved the defining
Eq.~(\ref{eq:defining_eq}) numerically by iteration, starting from a
Maxwell-Boltzmann distribution. The convergence of the iteration
process is very fast; there is almost no difference visible between
the first three iterations (see Fig.~\ref{fig:beta5}). To quantify the
difference between two subsequent iterations, we compute the ${\cal
  L}^1$ norm of $\Delta f(x)=f^{n+1}(x)-f^n(x)$. As an example, for
$\beta=3$ we find values of $\mathcal{O}(10^{-3})$ between the first
and second iteration, and $\mathcal{O}(10^{-9})$ between the second
and third iteration, respectively. The iterative solution of
Eq.~(\ref{eq:defining_eq}) is compared to the data from simulations
for several values of $\beta$ in Fig.~\ref{fig:velocity_dist_0.35}. No
deviations can be detected within the scatter of the data. We find
similar good agreement for {\bf all} values of $\beta.$

\section{Conclusion}

We have shown that a Brownian suspension of interacting particles,
subjected to random accelerations, exhibits strongly anomalous
velocity distributions. An event driven algorithm was generalised to
finite friction, allowing for large scale simulations of over 2
million particles. The simulations reveal velocity distributions which
are universal in the sense that they are largely independent of volume
fraction and collisions between the particles, and only depend on
damping rate and kick frequency through the ratio
$\beta=\gamma/f_{Dr}$. This has led us to consider a simplified one
particle model allowing for an analytical theory of the velocity
distribution, $f(v)$. The resulting integral equation reduces
trivially to the Maxwell-Boltzmann distribution for $\beta \to 0$. For
large $\beta$, we find a divergent distribution for small argument,
$f(0)\sim e^{\beta}$, a $1/v$ decay for intermediate $v$ and Gaussian
behavior for the largest argument. Hence there are no exponential
tails. In Refs.~\cite{Romanczuk2012,VanDenBroeck83} an exponential
tail was obtained for a damped particle kicked by white shot noise,
but in these works the kick size distribution was exponential, rather
than the Gaussian we use.  For intermediate $\beta$, the integral
equation for $f(x)$ is solved by iteration with very fast
convergence. For all $\beta$ we find excellent agreement between the
one particle theory and the simulations.

Power law velocity distributions are nontrivial solutions of the
unforced Boltzmann equation~\cite{BenNaimMachta2005}, where
dissipation is due to inelastic collisions and no damping with a
medium is considered.  In contrast, for elastic collisions as
considered here, the solution of the Boltzmann equation is of course
the Maxwell-Boltzmann distribution.  Hence the origin of the algebraic
decay of the velocity distribution observed in the present work is
distinct from that of Ref.~\cite{BenNaimMachta2005}.

Our approach can be generalised in several ways. 
Both the simulations as well as the analytical
theory can be generalised to other distributions for the kick
amplitudes and times. It also of interest to include dissipation in
the collisions in order to make closer contact with experiments on granular
media. Furthermore we plan to study directed motion, polar particles
and rotational degrees of freedom, modeling other swimmers.

\begin{acknowledgments}
We thank C. Heussinger, W. T. Kranz, M. M\"uller and M. Wardetzky for
useful discussions.  A.F. and A.Z. acknowledge support from DFG by FOR
1394.
\end{acknowledgments}

\bibliography{letter.bib}

\end{document}